\documentclass[conference]{IEEEtran}
\usepackage{upquote}
\usepackage{color}
\usepackage{graphicx}
\usepackage{tabularx}
\usepackage{lipsum}
\usepackage{booktabs}
\usepackage{amsmath}
\usepackage[square,sort,comma,numbers]{natbib}
\DeclareGraphicsExtensions{.pdf,.jpeg,.jpg, .png, .eps, .tiff}

\ifCLASSINFOpdf
\else
\fi
\begin{document}
%
\title{Analysis of Vacancy defects in Hybrid Graphene-Boron Nitride Armchair Nanoribbon based n-MOSFET at Ballistic Limit}

\author{\IEEEauthorblockN{Anuja Chanana, Santanu Mahaptra}
\IEEEauthorblockA{Nanoscale Device Research Laboratory,\\
Department of Electronic Systems Engineering,\\ 
Indian Institute of Science, Bangalore--560012, India\\
Email:anuja@dese.iisc.ernet.in, santanu@dese.iisc.ernet.in}
\and
\IEEEauthorblockN{Amretashis Sengupta}
\IEEEauthorblockA{School of VLSI Technology, \\
Indian Institute of Engineering Science and Technology,\\
Shibpur, Howrah-711103, India\\
Email: a.sengupta@vlsi.iiests.ac.in}}


%


\maketitle

\begin{abstract}
Here, we report the performance of vacancy affected supercell of a hybrid  Graphene-Boron  Nitride embedded armchair  nanoribbon  (a-GNR-BN) based n-MOSFET at its ballistic transport limit using  Non Equilibrium Green\textquotesingle s Function (NEGF) methodology. A supercell is made of the 3p configuration of armchair nanoribbon that is doped on the either side with 6 BN atoms and is also H-passivated. The type of vacancies studied are mono (B removal), di (B and N atom removal) and hole (removal of 6 atoms) formed all at the interface of carbon and BN atoms. Density Functional Theory (DFT) is employed to evaluate the material properties of this supercell like bandgap, effective mass and density of states (DOS). Further band gap and effective mass are utilized in self-consistent Poisson-Schrodinger calculator formalized using NEGF approach. For all the vacancy defects,  material properties show a decrease which is more significant for hole defects. This observation is consistent in the device characteristics as well where ON-current (I$_{ON}$) and Sub Threshold Slope (SS) shows the maximum increment for hole vacancy and increase is more significant becomes when the number of defects increase.
\end{abstract}


%
\IEEEpeerreviewmaketitle

\section{Introduction}
Graphene, the first 2 dimension material, to be mechanically exfoliated from its bulk counterpart \cite{novoselov2004electric} has drawn a tremendous interest in research domains. Because of its unusual bandstucture i.e Dirac cone \cite{PhysRev.71.622}, it serves as an ideal material to scrutinize various electronic \cite{RevModPhys.81.109}, mechanical \cite{lee2008measurement} and optical properties \cite{falkovsky2008optical}. Due to a semi metallic nature graphene sheets are inapplicable for logic applications \cite{schwierz2010graphene}. Till now many efforts are dedicated to open a band gap in graphene, one of which is lateral confinement of carriers through the formation of graphene nanoribbon (GNR) \cite{PhysRevLett.97.216803}. But the band gap diminishes to very low values with the increase in the nanoribbon width beyond 4 nm \cite{PhysRevLett.98.206805} and is also dependent on the chirality of the GNR. Embedding  Graphene  with  Boron  Nitride (nearly same lattice constant with graphene) is one of the effective ways of opening a band gap in gapless graphene \cite{seol2011bandgap,Tang201298,  noei2012computational, zhao2013electronic}. BN nanoribbons (BNNR) \cite{park2008energy} possess a higher band gap in comparison to GNR and embedding it with GNR can be effective in increasing the band gap of the nanoribbon \cite{tran2014large, ding2009electronic}. Such atomic layers of hybrid Graphene–-Boron Nitride have been synthesized experimentally \cite{ci2010atomic} and can be employed in future nanoelectronics.
\\In the present study we analyze the device performance characteristics at the ballistic limit of hybrid a-GNRBN supercell with various types of vacancy defects. The BN doping is symmetric with 6BN atoms on either side of the nanoribbon. Among all the 3 configurations of armchair nanoribbons namely 3p, 3p+1 and 3p+2, 3p has the maximum band gap and we have considered a supercell of 3p nanoribbon. The vacancy defects are monovacancy (single B atom removal), divacancy (one B and one N atom removal) and hole vacancy (removal of 2 B, 2 N and 2 C atoms) formed at the interface of graphene and BN nanoribbon and are distributed randomly. The band gap and effective mass of the pure and defected supercell is evaluated using the DFT simulations. Utilizing these material properties, we solve the self-consistent Poisson-Schrodinger equation under the NEGF formalism and thus evaluate the ballistic n-MOSFET device characteristics. Further, several output and transfer device characteristics such as I$_D$-V$_D$, I$_D$-V$_G$ and Sub-threshold Slope are studied. For a channel length of 10 nm, the transport is presumed to be purely ballistic.

\section{Methodology}
The supercell considered for the present study is formed by repeating the 3p configuration of hybrid a-GNR-BN consisting of total 42 atoms (30 GNR atoms and 12 BNNR  atoms), with 6 BN atoms on either side of the nanoribbon. The width of the nanoribbon is 5.05 nm and the length of the supercell is 3.23 nm. This particular configuration is found to have the maximum band gap according to the previous reports \cite{chanana2014performance}.  This particular size of the supercell consisting of 315 honeycombs, is chosen to study the maximal feasible effect of vacancy defects on the device performance. Figure 1(a) shows  supercell with 3 hole vacancy defects at the interface distributed randomly and are likely to be formed here \cite{PhysRevB.85.045422}. The monovacancy, divacancy and  hole vacancy defects at the interface are expanded in Figure 1(b), (c) and (d). The atoms are H-passivated so as to reduce the impact from edge states.\\ 
The materials properties are calculated using DFT-LDA simulations on Quantumwise ATK \cite{QumWS}. The Local Density Approximation (LDA) exchange correlation with Perdew-Zunger parametrization \cite{PhysRevB.23.5048} is employed for the present study. The basis set used is Double Zeta Polarized (DZP) having a mesh cut-off energy of 75 Hartree. Troullier-Martins type norm -conserving pseudopotential \cite{PhysRevB.43.1993} sets in ATK (NC-FHI [z=1] DZP for Hydrogen, NC-FHI [z=4] DZP for Carbon, NC-FHI [z=3] DZP for Boron and NC-FHI [z=5] DZP for Nitrogen) are used here. The Pulay-mixer algorithm is the iteration control parameter with the tolerance value of 10$^{-5}$. The  maximum number of iteration step is  considered as 100. The Monkhorst-Pack  k-grid  mesh for our simulations is 1x1x15 \cite{monkhorst1976special}. 

The material parameters are used in our in house NEGF simulator \cite{sengupta2013performance} to examine the device characteristics. In the NEGF formalism, the ballistic drain current is evaluated as 
\begin{multline}
I_{D}={\frac{4e}h}
\int_{-\infty}^\infty T(E)[f_{S}(E_{k,x}-\eta_{S})-f_{D}(E_{k,x}-\eta_{D})]dE
\label{equn_6}
\end{multline}
where e is the electronic charge, h is the Planck\textquotesingle s constant , f$_S$ and f$_D$ are the source and drain Fermi functions and $\eta$$_S$ and $\eta$$_D$ are the source and drain chemical potentials and
T(E) is the transmission matrix calculated as 
 \begin{equation}
T(E)=trace[A_{S}\Gamma_{D}]=trace[A_{D}\Gamma_{S}]
\label{equn_5}
\end{equation}
Here, A$_S$ and A$_D$ are the spectral densities and $\Gamma$$_S$ and $\Gamma$$_D$ are the broadening matrices evaluated as 
\begin{equation}
A_{S,D} = G(E)\Gamma_{S,D}G^\dag(E)
\label{equn_5}
\end{equation}

\begin{equation}
\Gamma_{S,D} = i[\Sigma_{S,D} - {\Sigma ^\dag }_{D,S}]
\label{equn_6}
\end{equation}
And further the Green\textquotesingle s function is calculated as 
\begin{equation}
G(E) = {[(EI - H - \Sigma_S- \Sigma_D]}
\label{equn_1}
\end{equation}

\begin{figure}[!t]
\centering
\includegraphics[width=3.5in]{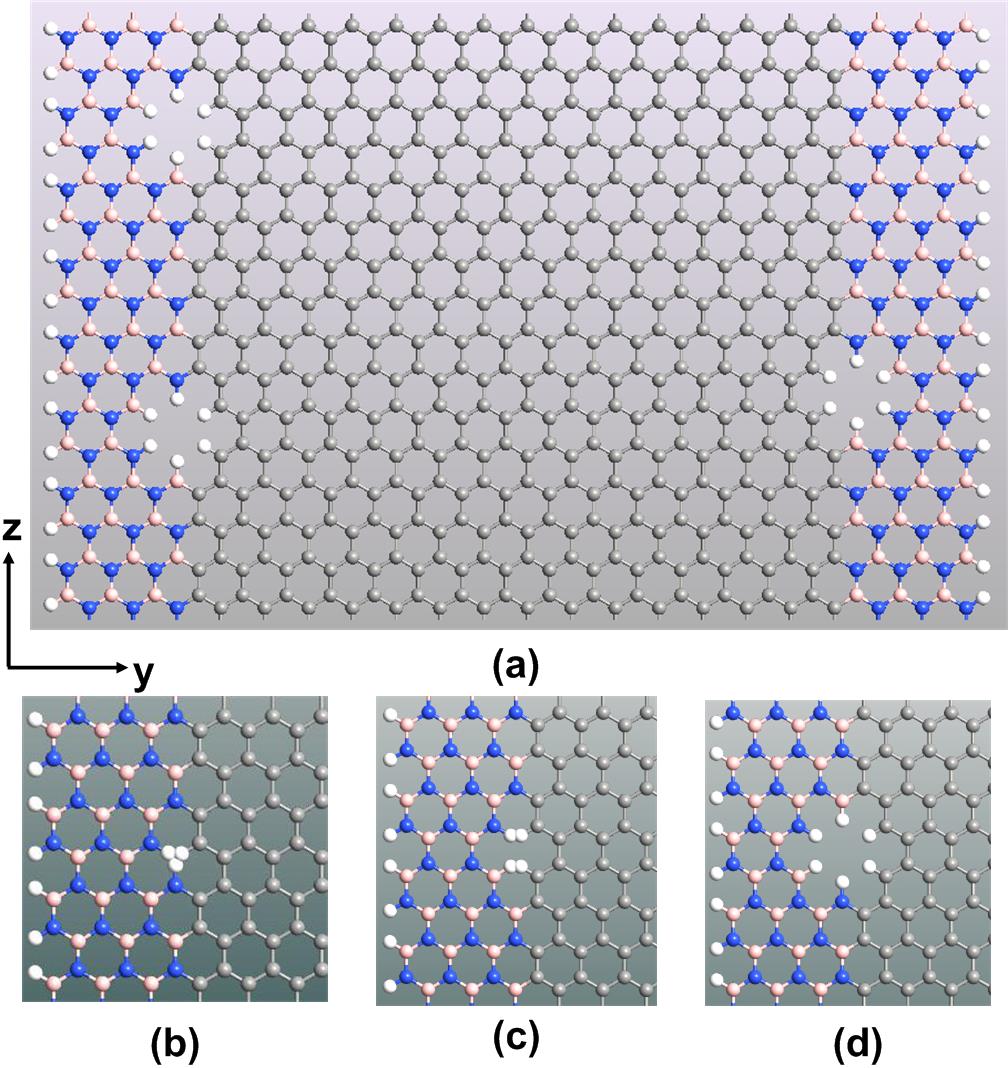}
\caption{(a) Supercell used in the present study. It is formed by repeating 8 times the unit cell 42aGNRBN. (b),(c) and (d) showing mono vacancy, di vacancy and hole defects. Hydrogen passivation is done to minimize contribution from edge states.}
\label{Figure 1}
\end{figure}

\section{Results and Discussions}
\begin{figure}[!t]
\centering
\includegraphics[width=3.5in]{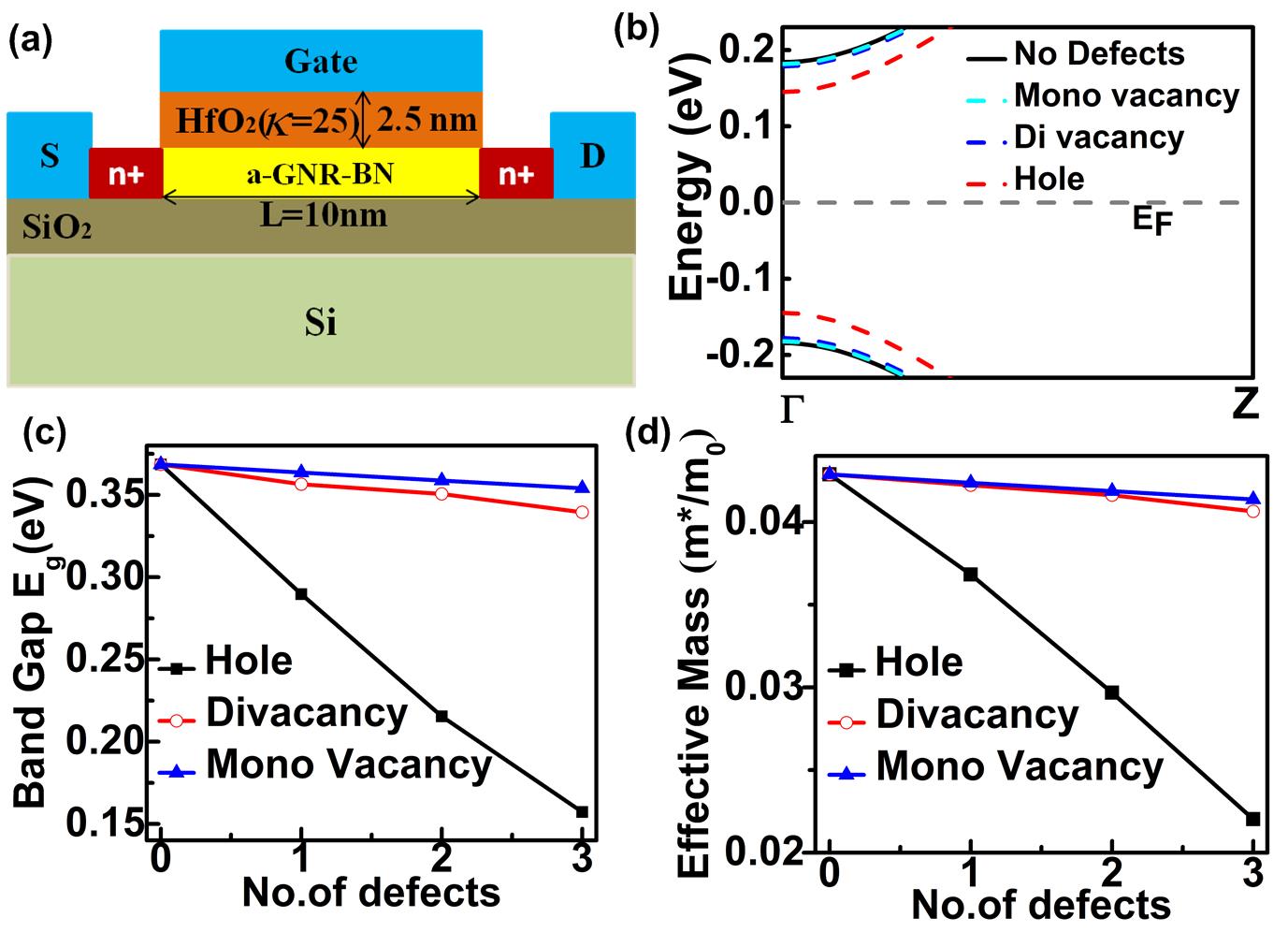}
\caption{(a)  Transistor schematic used for the device  calculations.  (b)  Comparison of band structures of pure and  defected supercell.(c) and (d) Band gap and Effective mass comparison with the increase in number of defects.}
\label{Figure 1}
\end{figure}

\begin{table}[!htbp]
\renewcommand{\arraystretch}{1.0}
\caption{Material parameters of pure and vacancy affected supercell evaluated using DFT (Number of defect is 
3 for each type of vacancy) }
\label{table_biax}
\centering
\begin{tabular}{l c c c r}
\hline
\bfseries Type of Vacancy defect  & \bfseries Band Gap (eV) & \bfseries Effective Mass (m$^*$/m$_0$) \\
\hline\hline
 Pure           & 0.369   & 0.0429       \\ \hline
        Mono vacancy   & 0.354   & 0.0413        \\ \hline
        Di vacancy     & 0.339   & 0.0407        \\ \hline
        Hole           & 0.157   & 0.022         \\ \hline
\hline
\end{tabular}
\end{table}   

Figure 2 (a) shows the transistor schematic of the n-MOSFET. Hybrid a-GNRBN supercell with vacancy defects is the 2D channel material. For the present analysis the channel length is assumed to be 10 nm and the width of the supercell is 5.05 nm. The gate dielectric is HfO$_2$ with a thickness of 2.5 nm and the channel is placed over SiO$_2$/Si substrate. Heavily doped n++ source/drain contact regions results in an effective alignment of contact Fermi Levels with the conduction and valence band of the channel material.\\
Figure 2(b) demonstrates a comparison of band structure of pure and vacancy affected super cell. The number of defects is 3 for each type. For both the pure and vacancy affected supercells, the bandstructure shows a direct nature at the gamma point of the Brillouin zone. We here observe that the band gap decrease as the number of removed atoms increase. The bandstructure lines nearly coincide with each other for monovacancy and divacancy defects with the pure ones, but for hole defects the decrease is substantial. Figure 2(b) and (c) shows the declining nature of both band gap and effective mass with the increase in the number of defects. For all types of vacancy defects in consideration we observe that the material properties show diminishing trend and it is maximum for the hole type vacancy defect. Table I provides the value of band gap and effective mass which is in a clear agreement with the trend depicted in Figure 2(b) and (c).\\

\begin{figure}[!t]
\centering
\includegraphics[width=3.5in]{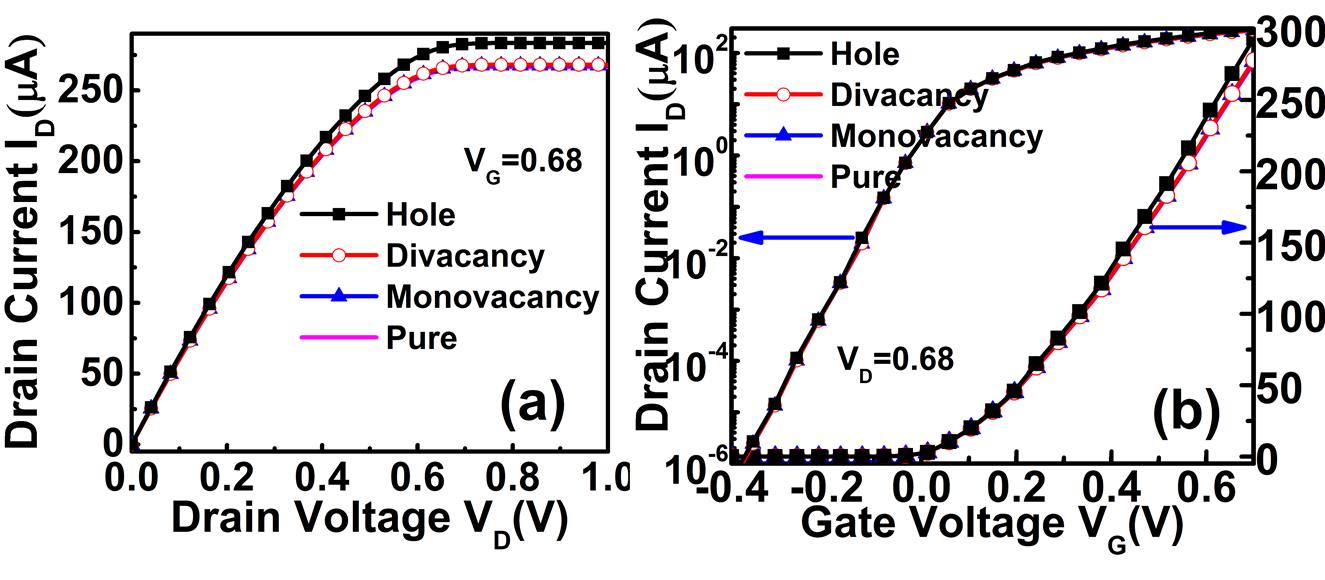}
\caption{(a) I$_D$-V$_D$ and (b) I$_D$-V$_G$ characteristics of pure and defected super cell. Defect density for each type of vacancy is 3.}
\label{Figure 1}
\end{figure}

\begin{figure}[!t]
\centering
\includegraphics[width=3.5in]{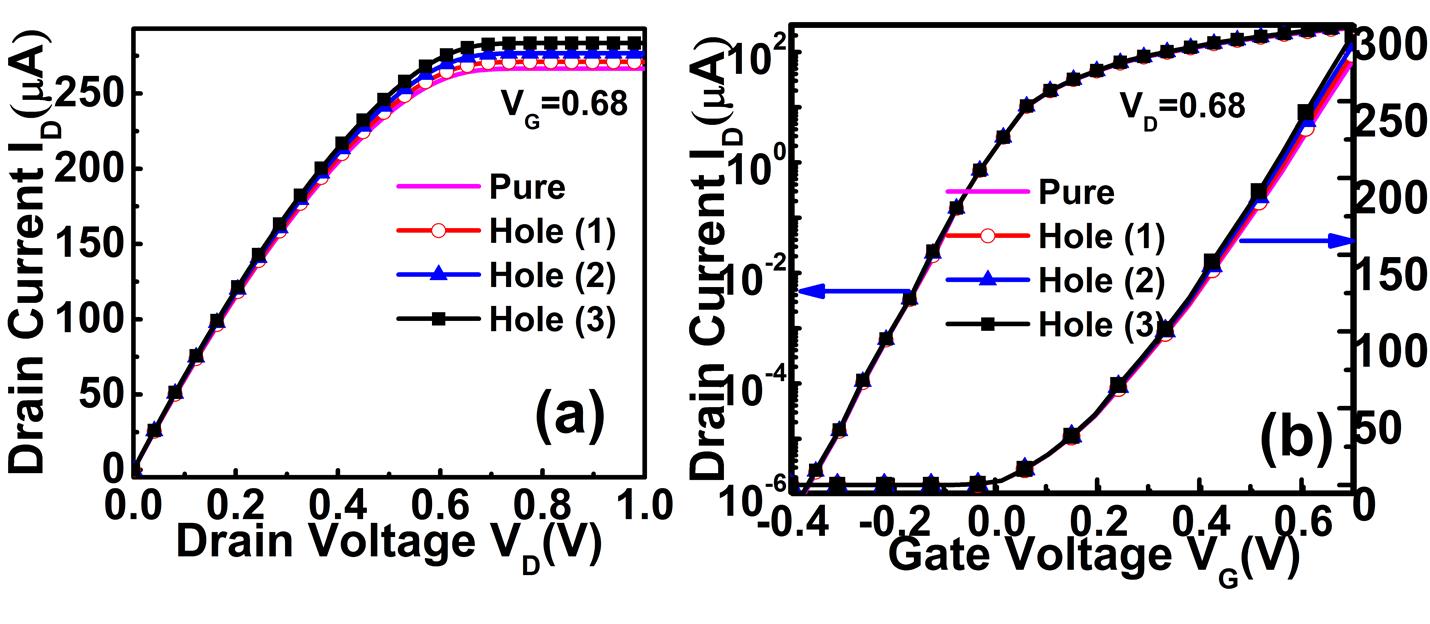}
\caption{(a) I$_D$-V$_D$ and (b) I$_D$-V$_G$ characteristics  hole defects from 1 to 3. }
\label{Figure 1}
\end{figure}
\begin{table}[!htbp]
\renewcommand{\arraystretch}{1.0}
\caption{n-MOSFET device parameters of pure and vacancy affected supercell calculated using NEGF formalism (No. of defect is 
3 for each type of vacancy) }
\label{table_biax}
\centering
\begin{tabular}{l c c r}
\hline
\bfseries Type of Vacancy defect  & \bfseries I$_{ON}$($\mu$A) & \bfseries Subthreshold Slope (mV/decade) \\ 
\hline\hline
 Pure             & 98.14 &  62.096      \\ \hline
        Mono vacancy     & 98.38 &  62.138      \\ \hline
        Di vacancy       & 98.5  &  62.189      \\ \hline
        Hole             & 101.6 &  62.622      \\ \hline
\hline
\end{tabular}
\end{table}   
               
Figure 3 and 4 depicts the NEGF simulated device characteristics of pure and defected supercells.   Number  of  defects 
considered is 3. The I$_D$--V$_D$ graph (Figure 3(a)) records the highest current for hole vacancy and nearly same current for pure and rest of the defected supercells. The ON current from pure to defected device varies in the range of 276-293  $\mu$A. The same effect is seen in the I$_D$-V$_G$ curve in Figure 3(b). Figure 4(a)-(b) shows the characteristics for increasing number of hole defects. A substantial increase in I$_D$ is seen when the number of hole defects is 3. For hole vacancy it  can be seen  for a defect density of 6.67\% - 20\% (1, 2 and 3 hole out of 15 at the interface) the current ranges from 277-293 $\mu$A. Table II shows the ON current values and Sub Threshold Slope of various vacancy defects. We here observe the highest ON current and SS for a hole vacancy in comparison to pure and other vacancies. Hereby, we conclude that the hole vacancy demonstrates the substantial effect both in material and device characteristics.

\section{Conclusion}
Here we study the effect of different types of vacancy defects in a hybrid a-GNRBN supercell and further its impact on the n-MOSFET device characteristics at the ballistic limit is examined. It is observed that the effective mass and the  band gap decrease with the increase of the vacancy defects and this change is very significant for hole defects. Similarly, the device characteristics illustrate a minimal change when the defect is mono or divacancy, but the change  is more considerable when the defect is a hole vacancy. The same trend can be observed with an increasing density of  number  of  hole  vacancy.  Hence, it can be noted that introduction of hole defects leads to a  significant  alterations in the material and device characteristics.

\section*{Acknowledgment}
The work of A. Sengupta was supported by the DST, Government of India, through the DST INSPIRE Faculty Award DST/INSPIRE/04/2013/000108. The work of A. Chanana and S. Mahapatra  was supported by the DST, Government of India, tunder grant 
no. SR/S3/EECE/0151/2012.

\bibliography{Reference_Vacancy}
\bibliographystyle{IEEEtran}



%

\end{document}